\documentclass[aip,pof,amsmath,amssymb]{revtex4-1}
\usepackage{bm}
\usepackage{graphicx}
\usepackage{epstopdf}
\usepackage{color}
\usepackage{cleveref}
\usepackage{amssymb,amsfonts,amsmath}
\DeclareFontFamily{U}{euc}{}
\DeclareFontShape{U}{euc}{m}{n}{<-6>eurm5<6-8>eurm7<8->eurm10}{}%
\DeclareSymbolFont{AMSc}{U}{euc}{m}{n} 
\DeclareMathSymbol{\umu}{\mathord}{AMSc}{"16} 

\begin{document}

\newcommand{\fdrive}{f_\text{drive}}
\newcommand{\rbead}{r_\text{B}}
\newcommand{\rdimer}{R_\text{p}}
\newcommand{\wi}{W \! i}
\newcommand{\degreesC}{\,^{\circ}{\rm C}}
\newcommand{\degrees}{\,^{\circ}}

\title{Fluid-Induced Propulsion of Rigid Particles in Wormlike Micellar Solutions}

\author{David~A.~Gagnon}
\author{Nathan~C.~Keim}
\author{Xiaoning~Shen}
\author{Paulo~E.~Arratia}
\email{parratia@seas.upenn.edu}
\affiliation{Department of Mechanical Engineering and Applied Mechanics, University~of~Pennsylvania, Philadelphia,~PA 19104}

\date{\today}

\begin{abstract}
In the absence of inertia, a reciprocal swimmer achieves no net motion in a viscous Newtonian fluid. Here, we investigate the ability of a reciprocally actuated particle to translate through a complex fluid that possesses a network using tracking methods and birefringence imaging. A geometrically polar particle, a rod with a bead on one end, is reciprocally rotated using magnetic fields. The particle is immersed in a wormlike micellar (WLM) solution that is known to be susceptible to the formation of shear bands and other localized structures due to shear-induced remodeling of its microstructure. Results show that the nonlinearities present in this WLM solution break time-reversal symmetry under certain conditions, and enable propulsion of an artificial ``swimmer.'' We find three regimes dependent on the Deborah number ($De$): net motion towards the bead-end of the particle at low $De$, net motion towards the rod-end of the particle at intermediate $De$, and no appreciable propulsion at high $De$. At low $De$, where the particle time-scale is longer then the fluid relaxation time, we believe that propulsion is caused by an imbalance in the fluid first normal stress differences between the two ends of the particle (bead and rod). At $De\sim1$, however, we observe the emergence of a region of network anisotropy near the rod using birefringence imaging. This anisotropy suggests alignment of the micellar network, which is ``locked in'' due to the shorter time-scale of the particle relative to the fluid.
\end{abstract}

\maketitle

\section{Introduction}

Many microorganisms, including single-cell eukaryotic protozoa (e.g.~spermatozoa\cite{Brennen:1977p2384,Gray:1955p2323,Gray:1955p2263}), prokaryotes (e.g.~bacteria \cite{Cohen2010}), and multi-cellular organisms (e.g.~nematodes \cite{Cohen2010,Gray:1964p2324}) know only linear viscous stresses. This is because viscous stresses that scale as $\mu U/L$ are much larger than nonlinear stresses from fluid inertia that scale as $\rho U^{2}$, where $U$ is a characteristic velocity, $L$ is a length-scale, and $\rho$ and $\mu$ are the fluid density and viscosity, respectively. For microorganisms swimming in Newtonian liquids like water, the ratio of inertial to viscous stresses, calculated using the Reynolds number $Re=\rho UL/\mu$, is very small ($Re \ll 0.1$) due to organisms' small length scale ($L < 10^{-4}$ m).  A remarkable property of Newtonian fluid flow at low $Re$ is time-reversibility, also known as kinematic reversibility.  This means that microorganisms moving in viscous environments, such as \textit{E. coli} swimming in water, can achieve net motion only from non-reciprocal kinematics that break this symmetry;\cite{Brennen:1977p2384,Lauga:2009ku,Lighthill1976} this restriction is also known as the ``scallop theorem.''\cite{Purcell1977} Microorganisms have developed different strategies to move at low $Re$, as seen in the rotating flagella of \textit{E. coli},\cite{Berg1993} the sinusoidal undulations of \textit{C. elegans},\cite{Korta:2007p2443,Shen:2011p8052,Sznitman2010Force,Sznitman2010Biophys} and the cilial beating of \textit{Paramecium}.\cite{Lauga:2009ku} While much work has revealed the details of propulsion at low $Re$ in Newtonian fluids,\cite{Lauga:2009ku} many natural environments encountered by microorganisms, such as bacterial films, human mucus and tissues, and soil, contain polymers and/or particles and are not Newtonian. Cervical fluid and gastric mucus, for example, have been shown to possess non-Newtonian behavior including viscoelasticity and rate dependent viscosity, and are successfully navigated by microscopic swimmers.\cite{Katz1978, Eriksen1998,Montecucco2001,Celli2009} Recently, theoretical and numerical studies\cite{Zhu:2012ht, Fu:2007p2306, Teran:2008p3896, Teran:2011p2301, Spagnolie2013} as well as experiments\cite{Liu:2011wk, Shen:2011p8052, Espinosa2013} have shown that viscoelasticity can significantly affect the ability of an organism to propel itself, including by modifying the swimmer's kinematics. Whether fluid elasticity enhances or hinders self-propulsion seems to depend on the type of kinematics employed by the organism, such as undulatory traveling waves or rotating helices, and its interactions with the fluid microstructure, such as polymer molecules and networks. 

As briefly mentioned above, many investigations have shown that the nonlinear rheological behavior (e.g. viscoelasticity) characteristic of many complex fluids can modify the swimming behavior of microorganisms. Another possibility is that complex fluids possessing nonlinear rheology and/or microstructure (e.g. polymer networks) can \textit{enable} rather than just modify propulsion.\cite{Lauga:2009p8056,Keim2012} Such ``fluid-induced'' propulsion at low $Re$ has been theoretically predicted for idealized viscoelastic fluids.\cite{Normand:2008p5115,Pak:2010p5116} By solving the Stokes equation along with the Oldroyd-B constitutive model for viscoelastic fluids, it was shown that propulsion at low $Re$ is possible for flapping surfaces,\cite{Normand:2008p5115,Pak:2010p5116} ``squirming'' of a sphere with surface oscillations,\cite{Lauga:2009p8056} and a cylinder with a reciprocal stroke but direction-dependent rates.\cite{Fu:2009p3840} A recent experimental investigation has indeed shown that the extra elastic stresses present in dilute polymeric (viscoelastic) solutions can break the constraint of kinematic reversibility at low $Re$ and lead to propulsion even for a reciprocally actuated ``swimmer.''\cite{Keim2012} This purely elastic, fluid-induced propulsion is not possible in simple, Newtonian fluids under the same conditions. 

The investigations discussed above focused on dilute viscoelastic solutions, in the sense that the fluid medium does not possess a network. But many fluids found in nature and the human body are not dilute and often possess microstructure that arises from interactions or cross-linking among polymer chains, for example. Many organisms are known to move, feed, and reproduce in highly structured fluids such wet soil,\cite{Juarez2010, Jung:2010p2342} human mucus,\cite{Fauci:2006p2439} and tissues.\cite{Harman2012} The interplay between the fluid's internal microstructure (e.g. polymer networks) and self-propulsion is critical to many biological processes such as reproduction,\cite{Fauci:2006p2439} bacterial infection,\cite{Josenhans2002} and bio-degradation in soil.\cite{Alexander1991} Structured fluids possessing networks such as surfactant or wormlike micellar (WLM) solutions can exhibit many fascinating phenomena under applied stress including shear banding\cite{Miller2007, Moller2008,Fardin2012} and even fracture.\cite{Rothstein2008, Akers2006,Gladden2007} Despite recent advances (briefly discussed below),\cite{Leshansky2009,Fu2010,Harman2012,Gagnon2013, Juarez2010,Teran:2011p2301} the effects of the fluid networks and microstructure on swimming at low $Re$ are still poorly understood. 

The effects of fluid microstructure on swimming have been previously studied in theory\cite{Leshansky2009,Fu2010} and in experiments.\cite{Gagnon2013, Juarez2010, Jung:2010p2342, Harman2012, Berg1979} A theoretical analysis using the Brinkman model to approximate heterogeneous, gel-like environments showed that the fluid microstructure can lead to an enhancement in propulsion speed.\cite{Leshansky2009} Note that the Brinkman model treats the heterogeneous media as static inclusions in a viscous fluid. Using a two-fluid model, which allows for both dynamic and stationary inclusions, Fu \textit{et al.}\cite{Fu2010} showed that the fluid network can enhance the propulsion of a infinite sheet when the microstructure is stiff and compressible. Experiments have also shown the fluid microstructure can significantly affect the motility behavior of living organisms. For example, \textit{Escherichia coli} can exhibit enhanced propulsion speeds in non-dilute polymeric solutions,\cite{Berg1979} \textit{Caenorhabditis elegans} can swim faster in polydisperse (size) wet granular networks,\cite{Juarez2010,Jung:2010p2342} and spirochetes in gelatin can exhibit four motility states that are highly dependent on gelatin concentration.\cite{Harman2012} More recently, an experimental investigation showed that the swimming speed of \textit{C. elegans} is enhanced in a concentrated polymer solution that supports the local alignment of polymer molecules.\cite{Gagnon2013} It was proposed that the swimmer's stroke aligns the polymer molecules, creating a local anisotropy in fluid mechanical response; crowding of polymers at these concentrations creates a history dependence favorable to undulatory swimming. Effectively, the swimmer actively modifies the local properties of the fluid with its motion.\cite{Gagnon2013}

In this paper, we investigate the ability of a reciprocally actuated particle to translate through a complex fluid that possesses a network. A geometrically polar particle, a rod with a bead on one end (Fig. 1a, top panel), is actuated (rotated) via an external magnetic field. The magnetic field oscillates in the form of a square wave, which causes the ends of the particle to repeatedly and reciprocally sweep through the same subtended arc (alternating clockwise and counter-clockwise). We place this particle in a wormlike micellar (WLM) solution, which is susceptible to the formation of shear bands and other localized flow-induced structures, wherein the fluid microstructure is far from equilibrium.\cite{Fardin2012} Results show that a reciprocally actuated rigid particle is indeed able to achieve net motion in a WLM solution (see Fig.~\ref{trajectories}), which indicates that kinematic reversibility has been broken.  The behavior of the particle in the WLM solution is highly dependent on the period of oscillation and shear rate relative to the characteristic time-scale of the fluid. When the actuation period is much slower than the time required for the micellar network to heal, we observe translation towards the end of the polar particle with the bead. When the period is approximately equal to the characteristic time-scale of the fluid and the average shear rate becomes large, we observe the alteration of fluid microstructure near the rod-end of the polar particle, and the propulsion reverses direction. At periods much shorter than the relaxation time, we observe no propulsion (see~Fig.~\ref{trajectoriesQuant}). Our results highlight the importance of the fluid microstructure in the present investigation. 

\section{Experimental Methods}
\begin{figure}
\begin{center}
\includegraphics[width=5.5in]{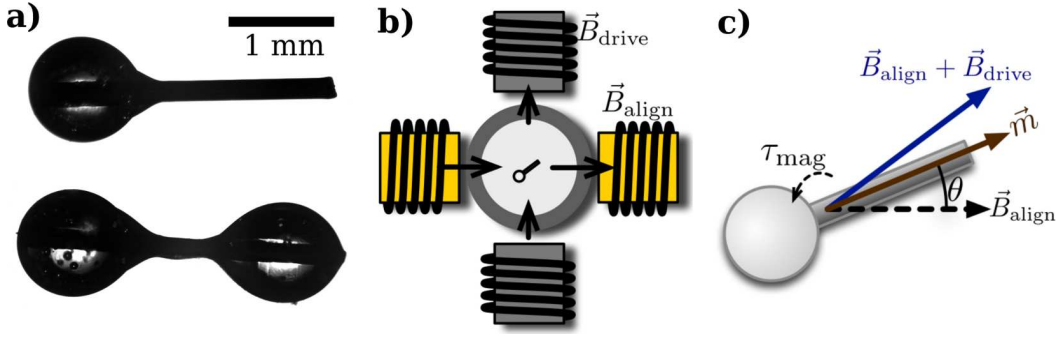}
\caption{{\textbf{(a)}} The two geometries investigated: (i) asymmetric and (ii) symmetric particle. {\textbf{(b)}} Apparatus consisting of one pair of aligning magnets orthogonal to a second pair of driving magnets. {\textbf{(c)}} Schematic of particle and apparatus.}
\label{setup}
\end{center}
\end{figure}

We investigate the possibility of fluid-induced propulsion in a WLM solution using two particle configurations: (i) asymmetric and (ii) symmetric, as shown in Fig.~\ref{setup}a. The asymmetric particle is fabricated from carbon steel wire of length $L \approx$~3~mm and radius $r_r \approx 115$~$\umu$m. An epoxy bead of radius $r_B \approx 500$~$\umu$m is placed at one end of this steel wire (Fig.~\ref{setup}a, top panel).  These ends will be referred to as the ``rod'' and ``bead'' respectively throughout the manuscript. The symmetric particle has two epoxy beads of radii $\rbead \approx 500$~$\umu$m, one on each end, as shown in Fig.~\ref{setup}a (bottom panel).

The particle is reciprocally actuated by four surrounding electromagnets, as shown in Fig.~\ref{setup}b. Two diametrically opposed electromagnets produce a constant field $\vec B_\text{align}$ while an orthogonal pair of magnets generate an AC field $\vec B_\text{drive}$. The AC field is computer controlled and driven in the shape of a square wave with a ramp time $t_r=0.05$~s, thus exerting a periodic torque $\tau_{mag}$ on the particle (Fig.~\ref{setup}c). Both the magnitude of $\vec B_\text{align}$ and amplitude of $\vec B_\text{drive}$ are on the order of $10^3$~G.\cite{Keim2012} The aligning magnitude and driving amplitude were kept constant and equal.

The magnetic fields actuating the particle are slightly inhomogeneous. While careful attention is paid to placing the particle in the center of the magnets, small discrepancies are unavoidable, and as a result, the particle can ``drift'' under the influence of the aligning magnets while in the absence of a driving magnetic field. In order to quantify the drift inherent in the apparatus, we performed tests to determine an upper bound for uncertainty in our experiments. This speed, $U \approx 1$ $\umu$m/s, serves a baseline for our results; speeds below this value should not be considered robust evidence of propulsion in this context.

The actuated particles are immersed in an aqueous solution of hexadecyl-trimethyl-ammonium bromide (CTAB; Sigma Aldrich, H5882) and sodium salicylate (NaSal; Sigma Aldrich, S2679) in deionized water.  CTAB, a common cationic surfactant, comprises a hydrophilic head and a hydrophobic tail.~\cite{Dreiss2007} These molecules self-assemble through a balance of weak polar repulsions and attractions into structures, known as micelles, at sufficient concentrations or when in the presence of salts or counter-ions.\cite{Dreiss2007, Miller2007} NaSal is known to promote the growth of cylindrical (hence wormlike) micelles.\cite{Dreiss2007} These micelles then entangle and form networks that break under sufficient stress and heal during relaxation.\cite{Miller2007} Here, wormlike micellar (WLM) solutions are prepared by slowly adding 130 mM CTAB to a solution of 130 mM NaSal and allowed to mix overnight. These are then degassed at room temperature using a vacuum chamber at 11.7~kPa for several hours until any bubbles have been removed. The WLM solution and the particle are placed in a glass container 50~mm in diameter and 30~mm tall, large enough to avoid wall effects.\cite{Keim2012}

\subsection{Particle Tracking}

The particles are tracked in space and time using computer vision methods. We create binary images by thresholding raw movies. The result is an image sequence which shows the filled contour of the swimmer (white, pixel intensity equal to 1) against a black background (pixel intensity equal to zero).  With these binary images, we can then compute image moments, which are weighted averages of pixel intensity.  The zeroth image moment reflects the area of the object, the first reflects the mean or centroid, and the second describes the axes of orientation. The particle velocity is obtained by differentiating the centroid data with respect to time. The particle orientation is obtained through taking the inverse tangent of a combination of second order image moments. The particle angular velocity $\dot{\theta}$ is calculated by differentiating the orientation data with respect to time. We then define shear rates at the rod-end and the bead-end of the particle. The instantaneous shear-rates are defined as
\begin{equation}\label{eq:g1}
\dot{\gamma}_R = \frac{\vert \dot{\theta} \vert R_{C,R}}{r_R} ; ~\dot{\gamma}_B = \frac{\vert \dot{\theta} \vert R_{C,B}}{r_B},
\end{equation}
while the cycle-averaged shear-rates are defined as
\begin{equation}\label{eq:g2}
\overline{\dot{\gamma}}_R = \frac{\sqrt{\langle \dot{\theta}^2 \rangle }R_{C,R}}{r_R}; ~\overline{\dot{\gamma}}_B = \frac{\sqrt{\langle \dot{\theta}^2 \rangle }R_{C,B}}{r_B}.
\end{equation}
In the above equations, brackets correspond to the mean over one cycle, $R_C$ denotes the distance to the center of rotation and the subscripts $R$ and $B$ refer to the rod-end and bead-end respectively, and $r_R$ and $r_B$ refer to the radii of the rod and bead respectively. 

Table~\ref{shearRate} summarizes these four shear rates for the particle at two values of Deborah number, defined as $De=2 \pi \lambda f$, where $\omega$ is the particle angular frequency, $f$ is the imposed frequency, and $\lambda$ is a characteristic fluid time-scale. (More details on $De$ and $\lambda$ are given below during our discussion of fluid rheology.) The table reflects two crucial features of our experiments. The first is the asymmetry of the particle shape: the disparity of $r_B$, $r_R$ and $R_{C,B}$ and $R_{C,R}$ results in $\dot \gamma_R$,~$\overline{\dot \gamma_R} \gg \dot \gamma_B$,~$\overline{\dot \gamma_B}$. The second is the dramatic expansion in the dynamic range of $\dot \gamma$ within a single cycle at high $De$, possibly due to shear-induced changes in fluid microstructure (Figs.~\ref{Biref1} and \ref{Biref2}).

\begin{table}
\begin{tabular}{| c | c | c |}
 \hline
  $De$ & 0.29 & 2.4 \\ \hline
  $\dot{\gamma}_{R}$ (s$^{-1}$) & 2.61 to 6.04 & 5.29 to 160 \\ \hline
  $\dot{\gamma}_{B}$ (s$^{-1}$ ) & 0.26 to 0.60 & 0.52 to 15.7 \\ \hline
  $\overline{\dot{\gamma}}_R$ (s$^{-1}$)  & 3.36 & 10.3 \\ \hline
  $\overline{\dot{\gamma}}_B$ (s$^{-1}$)  & 0.33 & 1.01 \\ \hline
  \end{tabular}
\caption{Typical shear rates estimated from high-speed video at $De=2 \pi \lambda f=0.29$ and $De=2.4$. Note the dramatic increase in the dynamic range of $\dot{\gamma}_R$.}
\label{shearRate}
\end{table} 

\subsection{Steady Rheology}

\begin{figure}
\begin{center}
\includegraphics[width=6.48in]{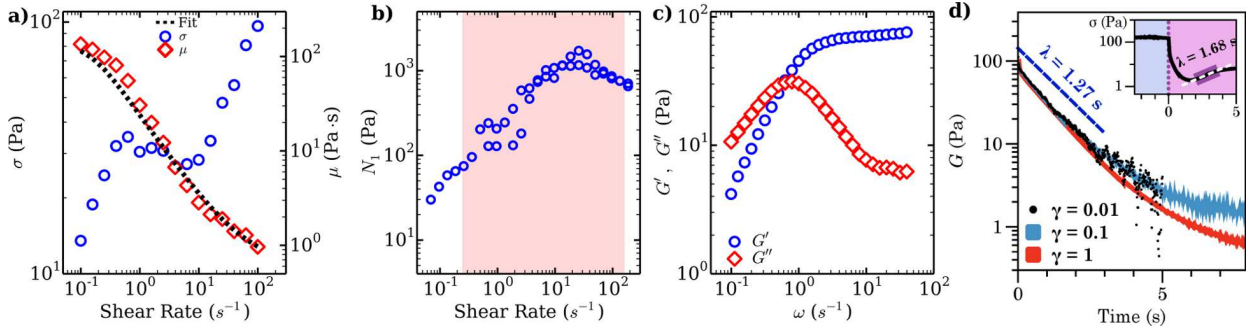}
\caption{{\bf{(a)}}~Shear viscosity $\mu$ and shear stress $\sigma$ as a function of shear rate $\dot{\gamma}$. The dashed line represents the Carreau model of shear-thinning. Note the stress plateau over $(0.65 \lesssim \dot{\gamma} \lesssim 10)$~s$^{-1}$. This plateau, where increasing shear rate does not result in increasing stress, is characteristic of fluids which support shear banding. {\bf{(b)}}~First normal stress difference $N_1$ as a function of shear rate. The shaded region denotes the range of shear rates in the vicinity of  the ends of the particle (see Table~\ref{shearRate}). {\bf{(c)}}~Storage ($G^{\prime}$) and loss ($G^{\prime \prime}$) moduli  as a function of angular frequency $\omega$ at strain amplitude $\gamma_0=0.01$. The inverse of the frequency where $G^{\prime}$ and $G^{\prime \prime}$ cross ($\omega_c \approx 0.64$~s$^{-1}$) defines a characteristic time-scale of the fluid ($\lambda \approx 1.56$~s). {\bf{(d)}}~Stress relaxation measurements at three different strains. The relaxation time in all three tests is consistent with $\lambda = 1.27$~s. Inset: stress relaxation measurement for step shear rate test using typical shear rates observed in our system ($100$~s$^{-1}$ followed by $0.05$~s$^{-1}$, $\gamma_0 \gg 1$). The fit suggests there is a characteristic time-scale, $\lambda_{rec} = 1.68$~s, which dominates the recovery of the WLM solution from large strains and high shear rate.  }
\label{rheology}
\end{center}
\end{figure}

The rheological properties of the wormlike micellar (WLM) solution, shown in Fig.~\ref{rheology}, are characterized using a cone-and-plate, strain-controlled rheometer (RFS3, TA Instruments). Figure~\ref{rheology}a shows the steady behavior of viscosity $\mu$ and shear stress $\sigma$ as a function of shear rate $\dot{\gamma}$. The observed plateau in shear stress between $0.65 \lesssim \dot{\gamma} \lesssim 10$~s$^{-1}$ is a well-known feature of WLM solutions and is often referred to as the unstable or plateau regime, in which the same shear stress can support multiple shear rates.\cite{Miller2007, Moller2008} This indicates that increasing or decreasing the applied shear stress through this unstable region will likely result in shear banding.\cite{Miller2007,Moller2008} This further implies that there can be local, abrupt variations in viscosity, as this solution exhibits strong shear thinning tendencies due to the aligning and breaking of WLM networks.\cite{Miller2007} The shear-thinning viscosity behavior of the WLM solution is characterized by fitting the rheological data to the Carreau-Yosuda model
\begin{equation}
\mu (\dot{\gamma})=\mu_{\infty} + (\mu_0 - \mu_{\infty}) (1 + (\lambda_{Cr} \dot{\gamma})^2)^{\frac{n-1}{2}},
\end{equation}
where $\mu (\dot{\gamma})$ is the fluid shear rate dependent viscosity, $\mu_0$ is the zero-shear viscosity, $\mu_{\infty}$ is the infinite-shear viscosity, and $n$ is the power law index of the fluid.\cite{Bird:1987p2262} The quantity $\lambda_{Cr}$ is a characteristic time-scale or inverse shear rate at which the fluid transition from Newtonian-like to power law behavior. This transition is often characterized by the Carreau number $Cr = \lambda_{Cr} \dot{\gamma} \approx 1$. If $Cr < 1$, then the fluid viscosity behaves Newtonian-like; if $Cr > 1$, then the fluid viscosity is shear thinning.\cite{Bird:1987p2262} Using a least squares fit, we find that $\mu_0 = 137$ Pa$\cdot$s, $\mu_{\infty} = 0.6$ Pa$\cdot$s, $\lambda_{Cr} = 6.9$ s, and $n = 0.1$. We can then estimate an upper bound for the Reynolds number, here defined as $Re = \rho \dot{\theta} L^2/2\mu_{\infty}$, where $\dot{\theta}$ is the angular velocity,  $L$ is the particle length, and $\rho$ and $\mu_{\infty}$ are the density and zero-shear viscosity of the solution respectively. We find the upper bound for $Re$ is approximately 0.08, which indicates that inertial effects may be neglected. 

WLM solutions are also known to develop significant normal stress differences. Fig.~\ref{rheology}b shows the first normal stress difference $N_1$ as a function of steady shear rate. The form of this curve is consistent with that of other WLM solutions.\cite{Anderson2006, Ober2011, Fardin2012}  The shaded region represents the range of shear rates experienced at the ends of the particle, suggesting that $N_1$ may play a role in our experiments.

\subsection{Unsteady Rheology and Characteristic Timescales}

We also perform unsteady rheological tests, in light of the time-varying $\dot \gamma$ in our experiments. Figure~\ref{rheology}c shows small amplitude oscillatory rheology ($\gamma_0=0.01$ or 1\%) for more than two decades of frequencies while Fig.~\ref{rheology}d shows stress relaxation measurements at three different strains ($\gamma_0 = 0.01$, $0.1$, and $1$). These tests indicate characteristic relaxation time-scales of the fluid in response to the magnetically-actuated particle in our experiments. We first measure the linear time-scale, $\lambda_{lin}$, which is a combination of the traditionally defined molecular reptation time ($\lambda_{r}$) and the equilibrium time-scale ($\lambda_{br}$) for the scission and reformation of micellar structures: $\lambda_{lin} = \sqrt{\lambda_r \lambda_{br}}$.\cite{Gladden2007, Walker2001, Rogers2012} This time-scale can be estimated from small-amplitude oscillatory tests (Fig.~\ref{rheology}c) by taking the inverse of the angular frequency $\omega$ where $G^{\prime}=G^{\prime \prime}$, where elasticity becomes dominant relative to viscous dissipation.\cite{Rogers2012} Using data from oscillatory rheology at a strain of $\gamma_0=0.01$, we estimate $\lambda_{lin}=1.56$~s. However, this time-scale represents the behavior in or near the linear regime of fluid response. It is not clear that a time-scale estimated from this linear regime will be relevant at large strains and in particular the high shear rates observed in our experiments (see Table~\ref{shearRate}).

In order to estimate fluid time-scales relevant to larger deformations, we examine data from stress relaxation (Fig.~\ref{rheology}d) and shear rate step tests (Fig.~\ref{rheology}d, inset). Figure~\ref{rheology}d shows stress relaxation tests at strains spanning two decades ($\gamma_0 = 0.01,0.1,$~and~$1$). We find that, for all the imposed strains, the stress relaxes at similar rates. We fit the relaxation data with a Maxwell model with a single time-scale of the form
\begin{equation}\label{eq:mts}
G(t) = G_0 e^{-t/ \lambda},
\end{equation}
where $G(t)$ is the shear modulus and $\lambda$ is the longest relaxation time of the fluid. We find a consistent relaxation time $\lambda$ for all strains of approximately 1.27~$\pm$~0.2~s. This suggests that the characteristic relaxation time $\lambda_{lin}$ from small-amplitude oscillatory rheology also dominates at strains two orders of magnitude larger than in those tests. 

Additionally, to examine the recovery of the WLM solution from high shear rates, we perform a shear rate step test (Fig.~\ref{rheology}d, inset), in which we impose a large, steady shear rate ($\dot{\gamma}=100$~s$^{-1}$) for 10~s ($\gamma_0 \gg 1$), immediately followed by a much lower steady shear rate ($\dot{\gamma}=0.05$~s$^{-1}$). This type of rheological test is dynamically similar to the experiments performed with the magnetic particle. We can then measure a characteristic time-scale for the fluid to recover from a high shear rate. The inset of Fig.~\ref{rheology}d shows the stress measured across this step in shear rate. We fit the data with an exponential function and obtain a time-scale of approximately 1.68 s. We note that this time-scale is similar in value to the time-scale measured from small amplitude oscillatory rheology (1.56~s) and to the time-scale measured from stress relaxation experiments (1.27~s). In the analysis below, we will use $\lambda \approx \lambda_{lin} = 1.56$~s, particularly in light of recent studies which suggest that the time-scale measured using small amplitude oscillatory rheology is relevant even at large strains.\cite{Rogers2012}

Now that we have estimated a value of $\lambda$, we can define two key dimensionless parameters, the Deborah number $De$ and the Weissenberg number $Wi$. The Deborah number is defined as $De=\lambda \omega = 2 \pi \lambda f$, where $\omega$ is the particle angular frequency and $f$ is the imposed frequency. Note that for a Newtonian fluid, $De=0$. The Weissenberg number is usually defined as the product of a characteristic strain-rate $\dot{\gamma}$ with the fluid relaxation time $\lambda$. Here, we can define different values of $Wi$ depending on the choice of characteristic shear rate (see~Eqs.~\ref{eq:g1} and \ref{eq:g2}). In this manuscript, we will mostly be concerned with the cycle-averaged Weissenberg number at the rod-end of the particle, defined as $\overline{Wi}_R =  \overline{\dot{\gamma}}_R \lambda$. 
  
In the context of our magnetically-actuated particle in a WLM solution, low $De$ ($De \ll 1$) indicates that the fluid has sufficient time to relax and reform its micellar network between particle reorientations, while high $De$ ($De \gg 1$) indicates that damaged micelles are unable to reform.  Most importantly, intermediate $De \sim 1$ indicates that the oscillation of the particle could potentially couple with the relaxation of the fluid, as each new oscillation occurs under the influence of the ``fading memory'' of the previous cycle, while not oscillating so quickly that the micellar structures are completely broken apart. The Weissenberg number $Wi$ can be viewed as the propensity of the particle's motion to deform the micellar structures, with high $Wi$ flows more likely to stretch or break the micellar network and to produce instabilities.\cite{Pakdel1996} 

\section{Experimental Results}

\begin{figure}
\begin{center}
\includegraphics[width=4in]{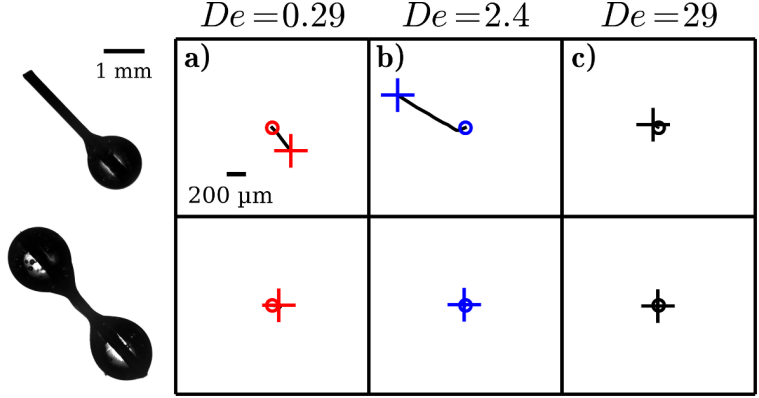}
\caption{Trajectories from $\circ$ to $+$ for asymmetric and symmetric particles actuated at three frequencies (corresponding to three values of $De$) for 60--80~s. The asymmetric particle exhibits robust propulsion at $De=0.29$ and $De=2.4$ towards the bead and rod respectively. No propulsion is observed for the symmetric particle or for the asymmetric particle at $De=29$; these small displacements are within the drift of the apparatus.}
\label{trajectories}
\end{center}
\end{figure}

For a Newtonian fluid at these low Reynolds numbers ($Re<0.08$), we would not expect a reciprocally actuated (rigid) particle, regardless of geometry, to achieve any appreciable net motion. A very different behavior, however, is found when the same particle is placed in a wormlike micellar (WLM) solution that is shear-thinning, viscoelastic, and prone to shear band formation. Despite the lack of inertia, we find that the reciprocally actuated particle is able to translate through the WLM medium (Fig. 3). 

Figure~\ref{trajectories} shows the displacement of both the asymmetric and symmetric particles immersed in a WLM solution at three values of $De$ ranging from 0.29 to 29. The particle displacement data is obtained by tracking the particle centroid using image analysis, as discussed in Sec.~II. We note that the particles shown in the far left of Fig.~\ref{trajectories} are oriented along the aligning field. At $De=0.29$, we observe that (i) the asymmetric particle moves parallel to the aligning field in the direction of the bead (Fig.~\ref{trajectories}a), and (ii) the symmetric particle translates negligibly. At $De=2.4$, we find that the asymmetric particle instead moves in the reverse direction, toward its rod (Fig.~\ref{trajectories}b), while the symmetric particle still shows no net motion. As the Deborah number is increased even further ($De=29$), we find that neither the asymmetric nor the symmetric particle exhibits propulsion. We note that neither the symmetric nor the asymmetric particle shows any net translation in Newtonian fluids under similar conditions (not shown). These observations indicate that the nonlinear rheology of the WLM solution, combined with an asymmetric geometry, break time-reversal symmetry and circumvent the so-called ``scallop theorem."\cite{Purcell1977} The top row of Fig.~\ref{trajectories} represents the three distinct regimes we find: (i) low $De$ or $De<1$, where the particle moves toward the bead; (ii) intermediate $De$ or $De \sim 1$, where the particle moves toward the rod, and (iii) high $De$ or $De>10$, where the particle does not appreciably translate. 

\begin{figure}
\begin{center}
\includegraphics[width=3in]{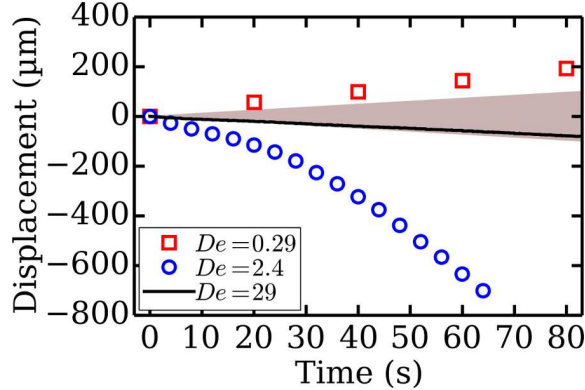}
\caption{Stroboscopic centroid displacement of the asymmetric particle tangent to the aligning field, as a function of time at three different values of $De = 2 \pi \lambda f$. The shaded region denotes the maximum displacement of the particle due to drift in the absence of a driving field.}
\label{trajectoriesQuant}
\end{center}
\end{figure}

Figure~\ref{trajectoriesQuant} shows the tangential displacement (parallel to the aligning field) of the asymmetric particle in Fig.~\ref{trajectories}, as a function of time. Each data point in Fig.~\ref{trajectoriesQuant} corresponds to one full cycle. We find that at $De=0.29$, the particle translates 200 $\umu$m in 4 cycles towards the bead ({\color{red}$\Box$}). In contrast, for $De=2.4$, the particle moves roughly 700 $\umu$m in 16 cycles ({\color{blue}$\circ$}), but this time in the direction of the rod. The third case, $De=29$ (black line), exhibits a small displacement, only 80 $\umu$m over 320 cycles, and is roughly equal to the maximum expected displacement due to drift for these experiments, denoted by the shaded area in Fig.~\ref{trajectories}. In what follows, we more closely examine these observations, discuss the transitions among the three observed regimes, and provide a potential mechanism to explain the behavior of this reciprocally-actuated asymmetric particle in WLM solutions.

\section{Discussion}
\label{sec:discussion}
\begin{figure}
\begin{center}
\includegraphics[width=5in]{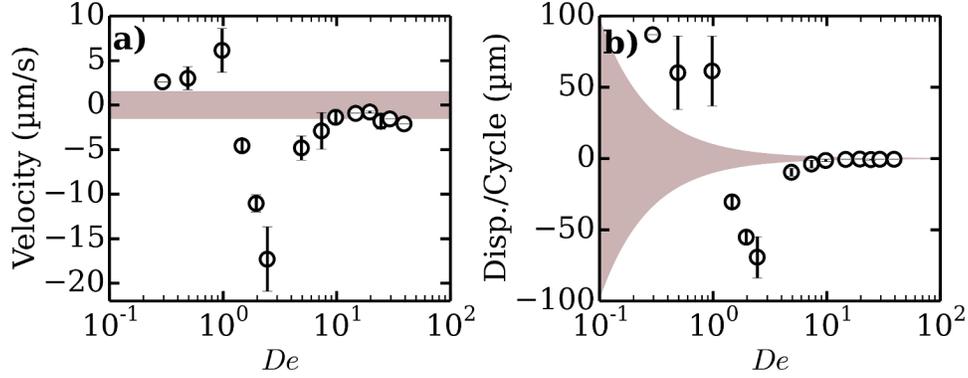}
\caption{{\bf{(a)}} Velocity and {\bf{(b)}} displacement per cycle as a function of $De$. A positive velocity and displacement per cycle indicate the particle moves tangential to the aligning field and towards the bead. A negative velocity represents motion in the opposite direction, towards the rod. Note the rapid transition from negative to positive velocities near $De=1$. The shaded area represents the maximum drift velocity ($\sim1$ $\umu$m/s) in Panel~(a) and the maximum displacement per cycle due to drift in Panel~(b), therefore providing an upper bound for uncertainty.}
\label{velocities}
\end{center}
\end{figure}

After identifying the three regimes discussed above, we would like to further examine the transition of the asymmetric particle from the regime with propulsion towards the bead for $De<1$ to the regime with propulsion towards the rod for $De \sim 1$, as well as provide a possible mechanism for the lack of propulsion for $De>10$.  Figure~\ref{velocities}a shows the particle propulsion velocity as a function of $De$ and Fig.~\ref{velocities}b shows the displacement per actuation cycle. A positive velocity and displacement per cycle indicate that the particle moves tangential to the aligning field and towards the bead. A negative velocity represents motion in the opposite direction, towards the rod. At low $De$ ($De<1.0$), we observe robust and repeatable propulsion, with large displacements per cycle ($\sim 75$ $\umu$m) in the direction of the bead. At intermediate values of $De$, that is, $1.0<De<10$, the particle moves rapidly towards the rod, again with large displacements per cycle ($\sim 75$ $\umu$m). There is a sharp peak at $De \approx 2.5$, and here the particle achieves speeds as high as 20 $\umu$m/s. Lastly, at high values of $De$ ($De>10$), we observe very little propulsion, with speeds roughly equal to the maximum drift speed for our apparatus in the direction of the rod and negligible displacements per cycle. 

The data discussed above clearly show that a reciprocally-actuated, rigid particle can achieve net motion in a WLM solution. Surprisingly, we find that the magnitude and direction of this net motion depend strongly on the Deborah number. One possible explanation for the observed particle propulsion at low $De$ is an imbalance in the first normal stress difference ($N_1$), normal to curved streamlines, between the two ends of the particle. Shear forces along curved streamlines are known to deform polymers non-uniformly, ultimately producing a force in the radial direction.\cite{Pakdel1996} Note that the rheology of the CTAB solution shown in Fig.~2b reveals that a significant first normal stress difference $N_1$ is present even at the shear rates produced by the particle. The rod end of the particle has a much larger shear rate (see Table 1) and greater curvature than the bead end, and this may produce an imbalance in $N_1$, and thus a force in the direction of the bead. Propulsion of this type has been previously observed for an asymmetric reciprocal dimer in a dilute polymeric (viscoelastic) solution.\cite{Keim2012} In other words, at low $De$, the interaction of the actuated particle and the CTAB solution seems to be similar to the response of a polar, rigid particle in a dilute polymeric solution. This type of polymeric solution is usually modeled using viscoelastic constitutive models such as Oldroyd-B and finite extensibility nonlinear elasticity (FENE).\cite{James:2009p3759}

The velocity and displacement data shown in Figure~\ref{velocities} demonstrate a clear transition, occurring in the vicinity of $De = \lambda \omega = 1$, which causes the particle to reverse direction. This indicates that the particle reverses direction when the period of driving is roughly equal to the characteristic time-scale $\lambda$ of the WLM solution. For our system, the characteristic time-scale $\lambda$ not only represents a bulk relaxation time of the fluid, but also a time-scale for the damaged micellar network to reform (see Fig.~\ref{rheology}c). At low $De$ ($De<1.0$), the micellar network has enough time to repair and heal all or nearly all damage caused by the last reorientation of the particle. At intermediate $De$, however, the actuated particle can encounter parts of the network that have not fully relaxed and/or healed since they were last disturbed. At large $De$ ($De>10$), the lack of propulsion suggests that the micellar structures are ripped apart by the motion of the particle and are unable to reform or heal. We can think of the micelles in this case (large $De$) as ``fluidized,'' removing the non-linear effects of the micellar network that would induce propulsion. Given that we observe a transition at $De \sim 1$, and negligible propulsion at $De > 10$, we believe that whether the micellar network has time to reform is crucial to the propulsion mechanism of the particle, and that the rate of healing relative to the frequency of motion dictates the direction the particle will translate. 

\begin{figure}
\begin{center}
\includegraphics[width=6in]{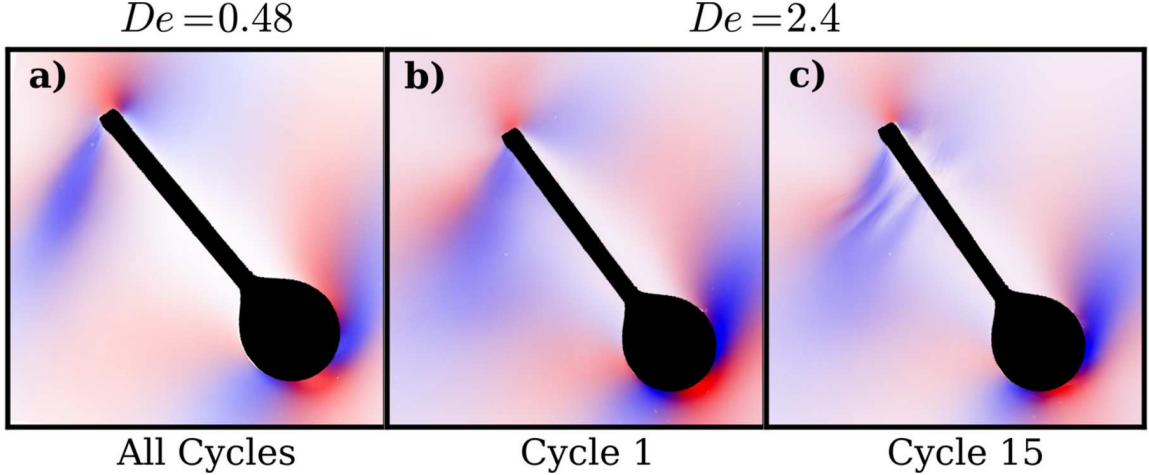}
\caption{(Color available online) Three snapshots of the particle nearly parallel to the aligning field and rotating clockwise. Two complementary polarization angles are shown in blue and red; blue is recorded with cross-polarizers parallel and orthogonal to the aligning magnetic field, while red shows light observed with cross-polarizers rotated 45 degrees relative to the blue axes. Intensity indicates the magnitude of birefringence detected. {\bf{(a)}}  Detection of birefringence at $De=0.48$; all cycles look identical. {\bf{(b)}} Detection of birefringence at $De=2.4$ after one cycle. Note the similarity with $De=0.48$. {\bf{(c)}} Detection of birefringence at $De=2.4$ after 15 cycles. Note the development of recurring striations indicating structure at the rod-end of the particle.}
\label{Biref1}
\end{center}
\end{figure}

\subsection{Imaging Fluid Anisotropy}

The data shown in Fig.~\ref{velocities} suggest that particle propulsion may be connected to the fluid microstructure. Hence, we examine the effects of the particle's motion using birefringence, a technique commonly used to study structure and fracture in WLM solutions.\cite{Gladden2007, Akers2006, Rothstein2008, Haward2012, Fardin2012}  High-speed video is taken at 150 frames per second while placing the WLM solution and particle between two cross-polarizers. While our camera images the experiment with a resolution of $\sim 10$~$\umu$m, the images probe the structure of the fluid at sub-micrometer length scales: Wherever the WLM network is completely isotropic, all light is extinguished by the second polarizer.  In regions where the network microstructure of the solution is anisotropic, the fluid is birefringent (with an index of refraction that depends on polarization direction) and can therefore rotate the polarization of transmitted light. This light is then partly admitted by the second polarizer and detected by the camera. In order to better capture the orientation of the network, we perform experiments with cross-polarizers at two different orientations to produce a composite image (Fig.~\ref{Biref1}a). The first orientation is parallel and orthogonal to the magnetic aligning field, and is represented in the composite image as blue. The second orientation places the cross-polarizers at 45 degrees relative to the ``blue axes" and is represented in the composite image as red.

Figure~\ref{Biref1} shows these composite images of birefringence for $De=0.48$ (low $De$) and $De=2.4$ (intermediate $De$) during the motion of the particle. Two images are shown at $De=2.4$, with Fig.~\ref{Biref1}b and Fig.~\ref{Biref1}c corresponding to the first and fifteenth cycles respectively.  In all three panels, birefringence is detected near the rod and the bead, indicating stress-induced anisotropy in the micellar network in those regions. The patterns detected at $De=0.48$ (Fig.~\ref{Biref1}a) and the first cycle at $De=2.4$ (Fig.~\ref{Biref1}b) look qualitatively similar. However, of particular interest are the striations (alternating patches of white and blue) formed along the rod in Fig.~\ref{Biref1}c. The development of these structures coincides with an increase in the cycle-averaged Weissenberg number near the rod-end of the particle, defined as $\overline{Wi}_R =  \overline{\dot{\gamma}}_R \lambda$, from $\mathcal{O}(1)$ at low $De$ to $\mathcal{O}(10)$ at intermediate $De$ (see Table~\ref{shearRate}). We note that the cycle-averaged Weissenberg number near the bead-end of the particle, defined as $\overline{Wi}_B =  \overline{\dot{\gamma}}_B \lambda$, similarly increases from $\mathcal{O}(0.1)$ to $\mathcal{O}(1)$. This rise in local shear rates and Weissenberg numbers is noteworthy because it has been shown that flow instabilities become more likely in polymeric and WLM solutions with increasing shear rate or $Wi$;\cite{Fardin2012,Pakdel1996} $Wi$ is often considered a measure of the non-linearity and stability of the fluid network. The striations in Figure~\ref{Biref1}c suggest that at sufficiently large values of Weissenberg number ($\overline{Wi}_R>10$), the oscillating particle is locally fracturing the micellar network and remodeling it into a form aligned with its rotation.\cite{Fardin2012, Gladden2007, Akers2006}

\begin{figure}
\begin{center}
\includegraphics[width=4in]{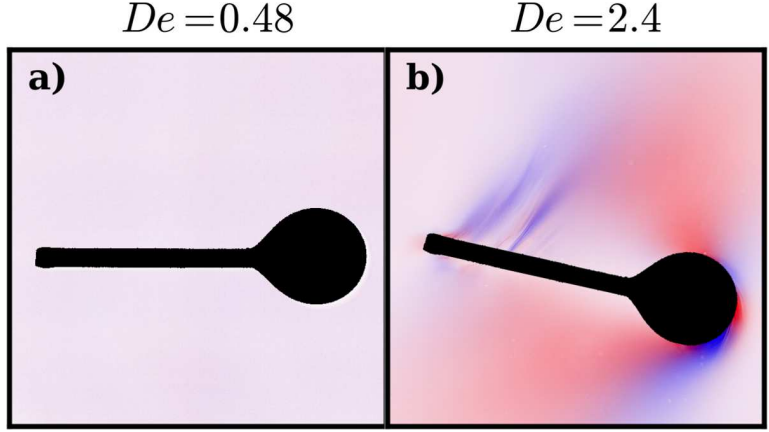}
\caption{(Color available online) {\bf{(a)}}  Absence of birefringence detected before the reversal in sign of the driving square wave at $De=0.48$. The fluid's network microstructure has relaxed to an isotropic equilibrium state. {\bf{(b)}} Birefringence detected before the reversal in sign of the driving square wave at $De=2.4$. Note that the birefringent regions formed during the rotation of the particle at intermediate $De$ persist for the entirety of each cycle.}
\label{Biref2}
\end{center}
\end{figure}

The observation that regions of altered microstructure develop during repeated shearing of the micellar networks should be reflected in the displacement data of the particle in Fig.~\ref{trajectoriesQuant}. In fact, the displacement per cycle is constant at $De=0.29$, but at $De=2.4$, there is an initial transient of $\sim 6$ cycles, after which the displacement per cycle grows. This transient behavior is paralleled by the development of striations in the birefringence images over many cycles (Fig.~\ref{Biref1}b,c). This requires both high Weissenberg number to generate the spatial pattern evident in the striations, and $De>1.0$ so that the micellar network cannot relax between reorientations, making the effects cumulative and the network anisotropy persistent. This persistence is evident in Figure~\ref{Biref2}, which shows the anisotropy of the micellar network in the frame immediately before the square wave switches sign to initiate a new half-cycle of motion. Figure~\ref{Biref2}a shows isotropy (absence of birefringence) in the network around the rod at $De=0.48$.  In stark contrast, Fig.~\ref{Biref2}b shows that at $De=2.4$, anisotropies in the network developed during the rotation of the particle are still present before the beginning of a new half-cycle of motion. In other words, the remodeling of the network that occurred during rotation due to $\overline{Wi}_R \gtrsim 10$ has accumulated and become ``locked in'' due to $De\gtrsim1$. 

\subsection{Propulsion at Intermediate $\boldsymbol{De}$}

In this section, we provide a possible mechanism for the particle propulsion at intermediate $De$. While the development of large-scale structures has been observed here using birefringence (Fig. 6c and Fig. 7b), we are unable to directly measure the fluid stresses or microstructure around the particle. Because of this limitation, we can only propose a mechanism using rheological data. At low $De$, we believe the particle translates as a result of an asymmetry in first normal stress difference $N_1$ produced at the ends of the particle. At intermediate $De$, we believe that this asymmetry is modified which leads to a reversal in particle propulsion direction.

First, our estimates show that the local shear rates imposed on the fluid by the particle are substantially larger at intermediate $De$ than at low $De$ (see Table 1). It is possible that, at the higher shear rates typical of $De\sim1$, the relative magnitudes of $N_1$ at the bead- and rod-end of the particle may be similar, at least when compared to relative magnitudes of $N_1$ at shear rates characteristic of low $De$ propulsion. For example, the rheological data in Fig. 2(b) shows that the values of $N_1$ increase by more than an order of magnitude between 0.1 and 5~s$^{-1}$; however, $N_1$ varies by less than a factor of two for shear rates between 5 and 100~s$^{-1}$. Since the effective cross-sectional area of the bead $A_B$ is larger than that of the rod $A_R$, somewhat similar magnitudes of $N_1$ at both ends of the particle may produce a larger radial force  ($\sim N_1 A$) near the bead, and thus a net imbalance in radial force in the direction of the rod if $N_{1,B} A_B > N_{1,R} A_R$. We stress that this admittedly simple explanation relies on steady rheology and lacks direct measurement of fluid stresses.

\begin{figure}
\begin{center}
\includegraphics[width=4.5in]{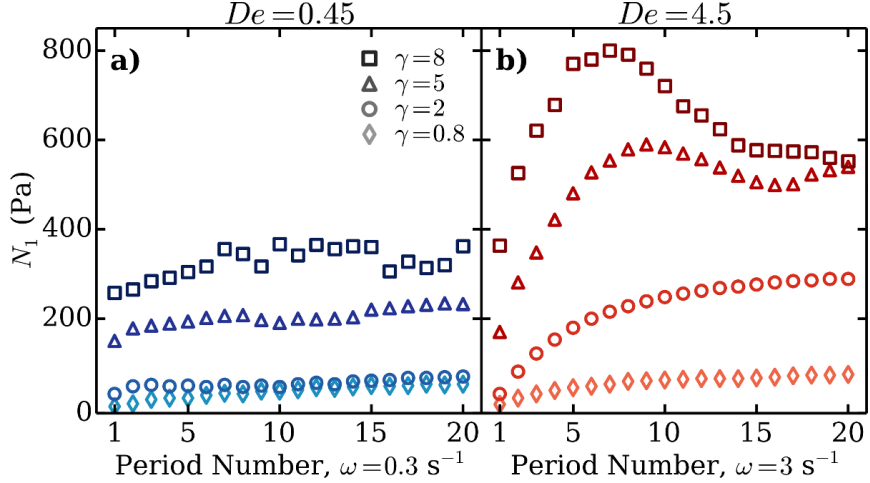}
\caption{{\bf{(a)}}~First normal stress difference $N_1$ as a function of period number under oscillatory shear at frequency $\omega = 0.3$~s$^{-1}$, which corresponds to $De = 0.45$. First normal stress difference at $De<1$ is largely independent of period number for each applied strain magnitude $\gamma = 0.8,~2,~5$, and $8$. {\bf{(b)}}~$N_1$ as a function of period number under oscillatory shear at frequency $\omega = 3$~s$^{-1}$, which corresponds to $De = 4.5$. First normal stress difference at $De\sim1$ increases initially with increasing number of periods, and for large strains ($\gamma = 5$ and $8$) decreases starting between period number 5 and 10. This decrease is then sustained through period number 20.}
\label{N1Oscil}
\end{center}
\end{figure}

Furthermore, the development of persistent, large-scale structures near the rod observed through birefringence could induce a non-monotonic relationship between $N_1$ and accumulated strain (or number of cycles) under repeated shearing. To examine this as a possibility, we measure $N_1$ of the WLM solution under oscillatory shear in a conventional cone-and-plate rheometer. Figure~\ref{N1Oscil} shows $N_1$ as a function of number of periods at two different frequencies ($\omega = 0.3$ and $3$~s$^{-1}$) and four different strains ($\gamma = 0.8,~2,~5$ and $8$). The frequencies chosen correspond to (low) $De = 0.45$ and (intermediate) $De = 4.5$, as shown in Figs. 8(a) and 8(b) respectively. We find that at low $De$ (Fig. 8a), $N_1$ is nearly constant with increasing period number at all strain magnitudes. In contrast, at intermediate $De$ (Fig. 8b), $N_1$ increases with period number initially for all four applied strains. Importantly, for the two largest imposed strains ($\gamma = 5$ and $8$) at intermediate $De$, we observe a decrease in the magnitude of $N_1$ starting between period number 5 and 10, which is sustained through period number 20. We note that each sample maintains constant values of $G^{\prime}$ and $G^{\prime \prime}$ throughout each measurement. 

The above observations hint that repeated shearing may modify the fluid stress near the particle after many cycles at $De \sim 1$, and perhaps, along with the larger shear rates characteristic of $De \sim 1$, contribute to the observed reversal in propulsion direction. We note that these rheological tests cannot precisely reproduce the conditions of the experiment. However, we feel this extension of our proposed low-$De$ mechanism to intermediate $De$, using measured particle shear rates and rheological data, provides some insight into a possible cause for the reversal of propulsion.  

\section{Conclusion}
In this paper, we have demonstrated that net motion or propulsion is possible for reciprocally actuated, rigid particles immersed in a wormlike micellar (WLM) solution even at low Reynolds number; no propulsion is observed with Newtonian fluids under similar conditions. We investigated this fluid-induced propulsion in WLM solutions using tracking methods as well as birefringence, which is used to obtain information on the fluid microstructure. We find different propulsion regimes for an asymmetric particle (Fig. 1a, top panel) depending on the Deborah number ($De=\lambda \omega = 2 \pi \lambda f$): net motion towards the bead at low $De$, net motion towards the rod at intermediate $De$, and no propulsion at high $De$ (Figs. 3-5). At low $De$, we believe propulsion is caused by an imbalance in the first normal stress differences between the two ends of the particle (bead and rod); the higher relative curvature of the streamlines near the rod when compared to those near the bead generates a net force in the direction of the bead. In this regime, the WLM solution has ample time to relax each time it is sheared, and therefore the response of the fluid to the particle is one characteristic of other viscoelastic fluids.\cite{Keim2012} However, at $De\sim1$, we observe network anisotropy near the rod using birefringence (Figs. 6c and 7b), which indicates alignment of the micellar structure.  This alignment is ``locked in,'' due to the shorter time-scale of driving relative to the fluid's characteristic time-scale. 

As shown in Figs. 4 and 5, we find a reversal in the particle's propulsion direction at $De \sim 1$. Because we are unable to directly measure fluid stresses near the particle, we propose a possible mechanism based on rheological data in order to provide some insight into the particle propulsion. In short, we believe that \emph{(i)} at larger shear rates (Table 1) the values of $N_1$ become similar at both ends of the polar particle (Fig. 2b), which means that the net radial force is predominately a function of the effective areas of the particle's bead and rod and \emph{(ii)} under repeated shearing at a rate faster than the fluid relaxation time, the magnitude of $N_1$ may be altered after many cycles (Fig. 8b). We note that the large-amplitude responses we report here may be sensitive to the composition of the WLM solution. Different surfactant or salt concentrations could result in qualitatively different fluid responses and propulsion regimes.

This work adds to our understanding of swimming in complex media, in particular in fluids with networks. Here, the interplay between structural relaxations of WLM solutions and reciprocal actuation results in two directions of propulsion, primarily distinguished by the time-scale of the stroke. This shows that the general principle of propulsion enabled by nonlinear rheology can in fact take many forms, depending on fluid microstructure, swimmer geometry, and stroke. By extension, these experiments also suggest a broad range of possibilities for artificial microswimmers in complex media for use in targeted drug delivery, lab-on-a-chip devices, and collective self-assembly.

\begin{acknowledgments}
We thank Peter Olmsted and Gabriel Juarez for helpful discussions. This work is supported by the Army Research Office through award W911NF-11-1-0488. DAG was supported by an NSF Graduate Research Fellowship.
\end{acknowledgments}

\bibliography{MagneticSwimmer_V15}

\end{document}